\documentclass[10pt,article,dvips,a4]{elsart}

\usepackage{amsmath}
\usepackage{times}
\usepackage{graphicx}
\usepackage{amssymb}
\usepackage{fullpage}
\usepackage{natbib}
\usepackage{color}
\usepackage{pifont}
\usepackage{floatflt}
\usepackage{anysize}

\empty

\newcommand{\exb}{$\vec{E}_r\times \vec{B}$ }
\definecolor{Light}{gray}{.80}
\definecolor{Dark}{gray}{.20}

\journal{Proc. of 12$^{th}$ ICPP}
\def\@abstractwidth{1.\textwidth}
\def\upup{\vspace{-10pt}}
\def\upupup{\vspace{-14pt}}

\begin{document}
\begin{frontmatter}
%
\parbox{6.3in}{\title{ELM triggering conditions for the integrated 
modeling of \\
H-mode plasmas}\vspace{-0.3in}
\author[SAIC]{A.~Y.~Pankin\thanksref{email}}
\author[LU]{G.~Bateman}
\author[MIT]{D.~P.~Brennan}
\author[SAIC]{D.~D.~Schnack}
\author[GA]{P.~B.~Snyder}
\author[JET]{I.~Voitsekhovitch}
\author[LU]{A.~H.~Kritz}
\author[FZK]{G.~Janeschitz}
\author[TXCORP]{S.~Kruger}
\author[SIIT]{T.~Onjun}
\author[Hydro]{G.~W.~Pacher}
\author[INRS]{H.~D.~Pacher}
\thanks[email]{E-mail: alexei.y.pankin@saic.com}%
\address[SAIC]{SAIC, 10260 Campus Point Dr., San Diego, CA. 92121}
\address[LU]{Lehigh University, Bethlehem, PA 18015, USA}
\address[MIT]{MIT, Cambridge, MA 02139}
\address[GA]{General Atomics, San Diego, CA 92186}
\address[JET]{JET-UKAEA, Culham Science Centre, UK}
\address[TXCORP]{Tech-X, Boulder, CO 80303}
\address[FZK]{FZK-PL-Fusion, Karlsruhe, Germany}
\address[SIIT]{SIIT, Klong Luang, Pathumthani 12121, Thailand}
\address[Hydro]{Hydro-Qu\'ebec,~Varennes,~Qu\'ebec,~Canada}
\address[INRS]{INRS, Qu\'ebec, Canada\vspace{-0.3in}}}

\begin{abstract}
\parbox{6.4in}{\noindent Recent advances in the integrated modeling of ELMy
H-mode plasmas are presented. A model for the H-mode pedestal
and for the triggering of ELMs predicts the height, width, and
shape of the H-mode pedestal and the frequency and width of ELMs.
Formation of the pedestal and the L-H transition is the
direct result of \exb flow shear suppression of anomalous
transport. The periodic ELM crashes are triggered by either the
ballooning or peeling MHD instabilities. The BALOO,
DCON, and ELITE ideal MHD stability codes are used to derive a new
parametric expression for the peeling-ballooning threshold. The
new dependence for the peeling-ballooning threshold is implemented
in the ASTRA transport code. Results of integrated modeling of
DIII-D like discharges are presented and compared with
experimental observations. The results from the ideal MHD
stability codes are compared with results from the resistive MHD
stability code NIMROD.}
\end{abstract}
\maketitle
\end{frontmatter}

\section{\upup Introduction}

Transport modeling of the edge of tokamak
plasmas is a challenging problem, because a wide range of time and
length scales need to be considered and many different elements of
physics are involved at the plasma edge. The physics topics that
are critically important for the plasma edge are the transition
from low- to high- confinement regime (L-H transition), H-mode
pedestal build up, anomalous and neoclassical transport at the
plasma edge, role of the \exb flow shear, triggering and dynamics
of the edge localized modes (ELMs). One of the
effective ways to test ideas for physics models is to combine them
within an integrative modeling code and compare the simulation
results with the experiments. Integrated modeling studies that
self-consistently take into account the effects of the plasma edge
have been developing recently~\cite{pacher04,lonnroth04b,pankin04}.
In particular, a new model for the H-mode pedestal and ELMs has been recently
developed by Pankin~{\it et al}~\cite{pankin04}. The model predicts
the height, width, and shape of the H-mode pedestal as well as the
frequency and width of ELMs. The model for the H-mode pedestal in
tokamak plasmas is based on flow shear reduction of anomalous
transport. The formation of the pedestal and the L-H transition
in this model are the direct result of \exb flow shear
suppression of transport. ELMs can be triggered either by 
ballooning or by peeling modes. The model for the
pedestal and ELMs has been used in a predictive integrated modeling
code to follow the time evolution of tokamak discharges from L-mode
through the transition from L-mode to H-mode, with the formation of
the H-mode pedestal, and, subsequently, the triggering of ELMs.
The model for the H-mode pedestal and ELMs~\cite{pankin04} is
advanced in this paper. The ELM triggering conditions are studied
with the MHD stability codes BALOO~\cite{miller87},
DCON~\cite{dcon}, and ELITE~\cite{wilson02}. These MHD instability
codes are used to compute the combined peeling-ballooning
threshold, which are then used to derive fitting expressions that
are included in the model. Using these MHD instability enhances
the model and extends the level of its applicability. 
The improved stability criterion model is tested in the integrated
modeling code ASTRA. Cases with low and high
triangularities are considered.\upupup

\section{Peeling-ballooning stability analysis\upup}\label{s-stable}

Plasmas with high triangularity, $\delta=0.6$, and low
triangularity, $\delta=0.2$, are considered. Other plasma
parameters are held fixed in the reference cases: the minor radius
$a=0.63$~m; major radius $R = 1.69$~m; toroidal magnetic field $B_T
= 2.0$~T; plasma current $I = 1.54$~MA; elongation $\kappa=1.78$;
central plasma density $n_e(0)=4.7\times 10^{19}$~m$^{-3}$; and
central ion end electron temperatures $T_{e,i}=4$~kEV. The TOQ
equilibrium code~\cite{miller87} is used to generate a set of
equilibria that covers the range of transport simulations for the
plasma parameters given above. As long as the plasma geometry, 
toroidal magnetic field, and total plasma current are fixed
in the transport simulations, ELMs are controlled by only the pressure
gradient and bootstrap current.  
The shape of the electron density profile is kept unchanged in all scans; the
electron density at the top of the pedestal is set to satisfy the
dependence $n_{\rm ped}=0.71 \left<n_e\right>$, which is observed
in experiments. In the density scan, the entire density
profile is scaled and in the
temperature scan, the central temperature is kept fixed, while the pedestal
temperature is changed. Both the bootstrap current and
pressure gradient are changed in the density and temperature scans. 
The density scan provides more control of the bootstrap
current, while the temperature scan provides more control of the
normalized pressure gradient, $\alpha$, which is defined in this study as
$\alpha =-\left({\mu_0}/{2 \pi^2}\right)\left({\partial p}/{\partial
\psi}\right)\left({\partial V}/{\partial \psi}\right) \left({V}/{2\pi^2
R}\right)^{1/2}$, where $V$ is the plasma volume and $\psi$ is the poloidal flux.
These equilibria are used in the BALOO, DCON, and ELITE codes to validate the
peeling-ballooning stability criteria in the limits of different toroidal mode numbers.
The BALOO code~\cite{miller87} is an infinite mode number ballooning
stability code developed at General Atomics. The ideal MHD DCON 
code~\cite{dcon} is suitable for the stability analysis of low toroidal number 
ballooning and peeling modes, and the
ELITE code~\cite{wilson02} works well for the analysis of intermediate and 
high mode numbers. 
Since these codes are complementary,
they can be used together to compute the stability criteria. 
\begin{figure}  \centering
\begin{tabular}{ll}
{\includegraphics[width=7.cm]{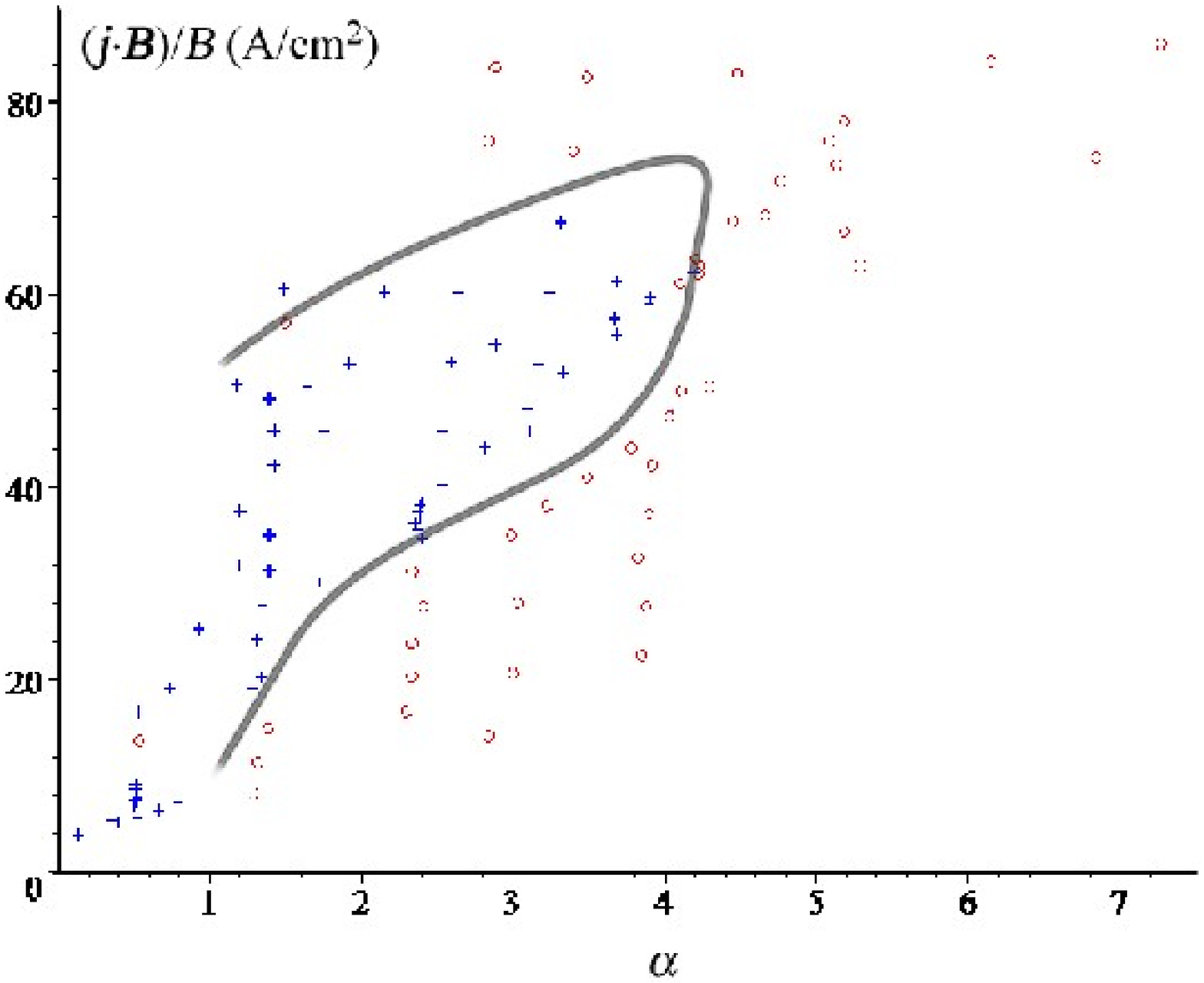}}&
{\includegraphics[width=7.cm]{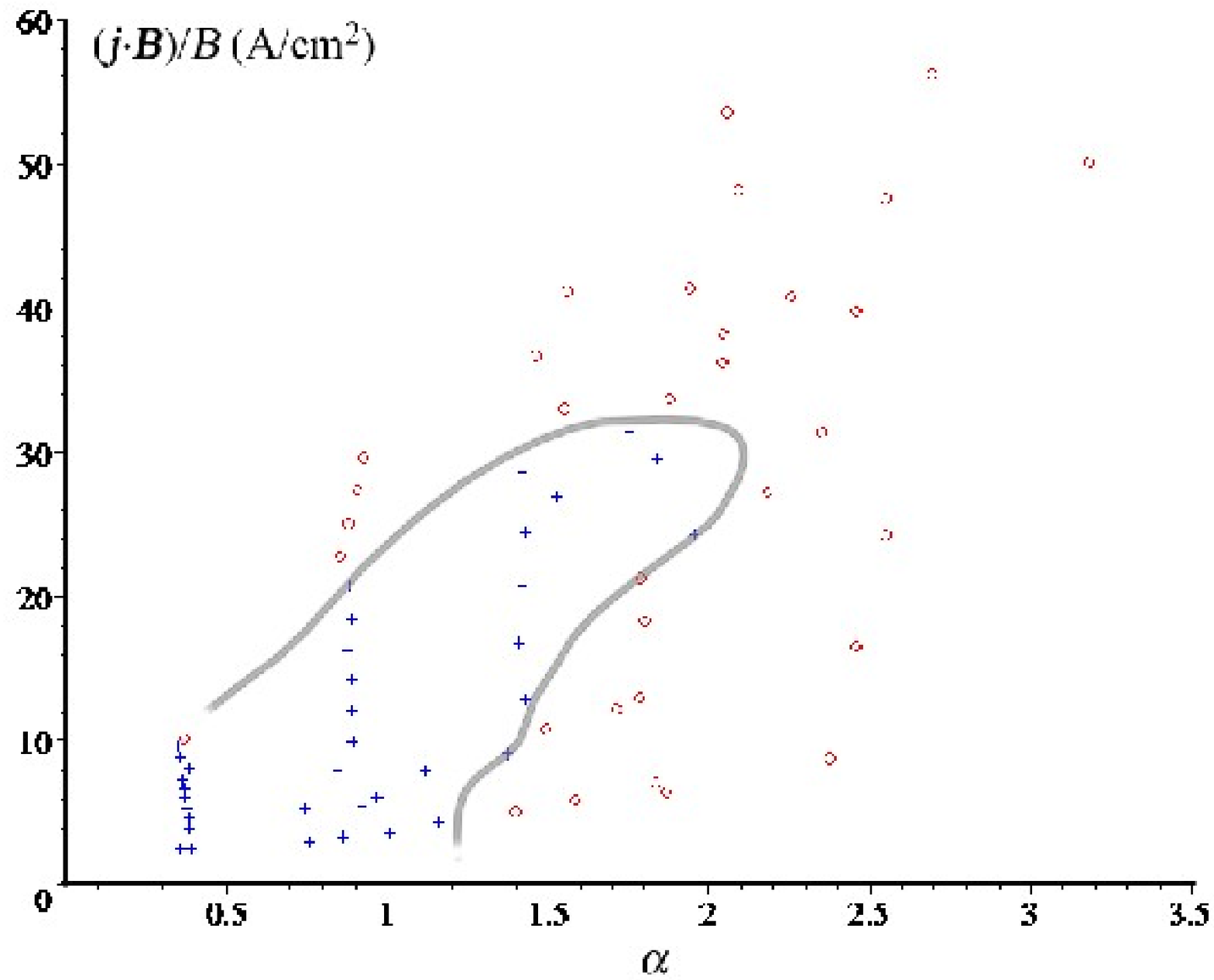}}
\end{tabular}
\renewcommand{\baselinestretch}{1.01}\normalsize
\caption{ELM stability diagrams for discharges with (a) high ($\delta=0.6$) and
(b) low ($\delta=0.2$) triangularity. Solid curve separates stable and unstable regions. The `+'
symbol on the diagrams corresponds to the cases that are tested with the MHD stability codes appear
to be stable; the 'o' symbol on the diagram correspond to the cases that appear to be unstable.
\label{f-stab}}
\end{figure}

The results of the stability analysis are shown in Fig.~\ref{f-stab}.
The high triangularity discharge has a larger stable region than the low
triangularity discharge, which is consistent with experimental results and other MHD stability analysis~\cite{snyder04}. 
In particular, the higher triangularity
discharges have a larger second stability region, which is also
consistent with the conclusion that higher triangularity discharges can more
easily access the second ballooning stability 
region of parameter space~\cite{onjun04b}.
The peeling-ballooning threshold shown in Fig.~\ref{f-stab} 
is parameterized using fifth order polynomials, which are implemented 
in the ASTRA transport code
and used as the criteria to trigger ELM crashes in the transport simulations.
\upupup%

\section{Results of integrated transport simulations\upup}\label{s-scans}

\begin{figure}  \centering
\includegraphics[width=9.5cm]{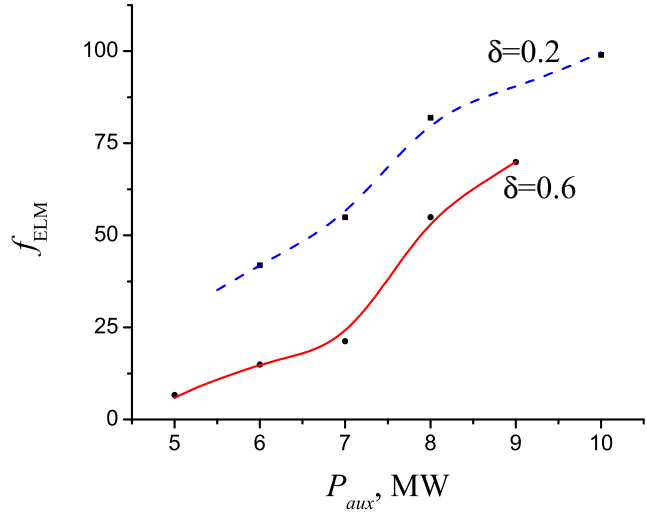}\upup
\renewcommand{\baselinestretch}{1.01}\normalsize
\caption{The frequency of ELM crashes as function of the auxiliary
heating power for discharges with
low ($\delta=0.2$) and high ($\delta=0.6)$ triangularities.}
\label{f-freq}
\end{figure}

A reference scenario for ASTRA simulations is based on typical DIII-D
geometry, using the parameters given at the beginning of previous Section. In
addition, the electron, ion, and impurity density profiles,
toroidal rotation velocity, $Z_{\rm eff}$, the current density driven by the
neutral beam injection (NBI) heating, and
the auxiliary heating power deposited to electrons and ions,
which are obtained from an analysis simulation
of experimental data, are prescribed and fixed in form.
For both the lower and higher triangularity discharges,
the auxiliary heating
power is varied from 3.5~MW to 7.0~MW in a series of simulations.
The ELM frequencies as a function of auxiliary heating power for discharges
with higher and lower triangularities are shown in
Fig.~\ref{f-freq}.
It can be seen that the ELM frequency increases with the heating power in
the simulations, which is consistent with experimental observations in
H-mode plasmas with type I ELMs.
The change of the slope of the ELM frequency as a function of heating
power, shown in Fig.~\ref{f-freq}, can be
explained by the different scenarios that are followed 
for low and high auxiliary heating discharges.
In the ASTRA simulations, the discharges with low auxiliary heating 
(below 7~MW) have ELM crashes that are triggered by a 
ballooning instability in the second stability limit, while
the discharges with high heating power (above 7~MW) are triggered
by a ballooning instability in the first stability limit.\upupup

\section{Summary\upup}\label{s-conclusions}

An improved model is introduced
for H-mode pedestal and ELMs~\cite{pankin04}. 
A parameterized peeling-ballooning stability criterion is implemented
in the model, based on detailed MHD analyses with the BALOO,
DCON, and ELITE codes. 
Two different scenarios for ELM crashes in DIII-D
discharges are shown. 
For the scenario with lower auxiliary heating power, ELMs are mostly
caused by the ballooning instability in the second stability limit. 
For the scenario with higher auxiliary heating power (above 7~MW), 
ELMs might be caused by the ballooning
instability in the first stability limit. 
Such ELM crashes are much less violent and more frequent. 
In general, the frequency of ELMs increases with the auxiliary
heating power (as shown in Fig.~\ref{f-freq}), 
which is consistent with the experimental observations.
The frequency of ELMs also depends on the plasma shaping.
In particular, the dependence on the triangularity is studied in this paper.
It is found that higher triangularity
discharges have a larger stability region than lower triangularity discharges
(compare Figs.~\ref{f-stab}~(a) and (b)). 
This observation is consistent with other MHD stability 
analysis~\cite{snyder04,onjun04b}. As result, 
ELMs in lower triangularity discharges are much more frequent than ELMs in
higher triangularity discharges (as shown in Fig.~\ref{f-freq}).

In conclusion, it is clear that additional MHD stability studies are
required. 
In this paper, ideal MHD stability codes are used, while
resistivity and two-fluid effects are expected to be important. 
A preliminary study with
the resistive MHD NIMROD~\cite{nimrod} code is under way. 
In order to verify the results
obtained with the MHD ideal stability code, 
a robust vacuum code should be used
together with the NIMROD code which 
will be done in future studies.\upup

\renewcommand{\baselinestretch}{1.65}\normalsize
\vspace{0.1in}
\vspace{-0.06in}
\renewcommand{\baselinestretch}{1.45}\normalsize
\bibliographystyle{revtex}
\renewcommand{\baselinestretch}{1.01}

\end{document}